\documentclass[conference, twocolumn, a4paper]{IEEEtran}

\IEEEoverridecommandlockouts

\usepackage{amsmath,amssymb,amsfonts}
\usepackage{graphicx}               
\usepackage{textcomp}
\usepackage{xcolor}
\usepackage[nolist,nohyperlinks]{acronym}
\usepackage{lipsum}
\usepackage{booktabs}
\usepackage{algorithm2e}
\usepackage{float}
\usepackage{nicefrac}
\usepackage{thmtools}
\usepackage{enumerate}
\usepackage[caption=false,font=footnotesize]{subfig}
\usepackage[hidelinks]{hyperref}    

\graphicspath{{figs/}}
\DeclareGraphicsExtensions{.pdf,.png,.jpg} 

\def\mathlette#1#2{{\mathchoice{\mbox{#1$\displaystyle #2$}}%
                           {\mbox{#1$\textstyle #2$}}%
                           {\mbox{#1$\scriptstyle #2$}}%
                           {\mbox{#1$\scriptscriptstyle #2$}}}} 
\renewcommand{\Vec}[1]{\mathlette{\boldmath}{#1}}

\begin{document}

\title{Adaptive RIS Control for Mobile mmWave NLoS Communication Using Single-Bit Feedback}
\author{Hamed Radpour$^\ast$, Markus Hofer$^\ast$, Thomas Zemen$^\ast$\\
	$^\ast$AIT Austrian Institute of Technology, Vienna, Austria\\
  	Email: hamed.radpour@ait.ac.at}

\maketitle

\begin{acronym} 
\setlength{\itemsep}{-0.63\parsep}
\acro{3GPP}{3rd Generation Partnership Project}
\acro{ACF}{auto correlation function}
\acro{ADC}{analog-to-digital converter}
\acro{BEM}{basis-expansion model}
\acro{CDF}{cumulative distribution function}
\acro{CE}{complex exponential}
\acro{CIR}{channel impulse response}
\acro{CP}[CP]{cyclic prefix}
\acro{CTF}{channel transfer function}
\acro{D2D}{device-to-device}
\acro{DAC}{digital-to-analog converter}
\acro{DC}{direct current}
\acro{DOD}{direction of departure}
\acro{DOA}{direction of arrival}
\acro{DPS}{discrete prolate spheroidal}
\acro{DPSWF}{discrete prolate spheroidal wave function}
\acro{DSD}{Doppler spectral density}
\acro{DSP}{digital signal processor}
\acro{DR}{dynamic range}
\acro{ETSI}{European Telecommunications Standards Institute}
\acro{FIFO}{first-input first-output}
\acro{FFT}{fast Fourier transform}
\acro{FP}{fixed-point}
\acro{FPGA}{field programmable gate array}
\acro{GSCM}{geometry-based stochastic channel model}
\acro{GSM}{global system for mobile communications}
\acro{GPS}{global positioning system}
\acro{HPBW}{half power beam width}
\acro{ICI}[ICI]{inter-carrier interference}
\acro{IDFT}{inverse discrete Fourier transform}
\acro{IF}{intermediate frequency}
\acro{IFFT}{inverse fast Fourier transform}
\acro{ISI}{inter-symbol interference}
\acro{ITS}{intelligent transportation system}
\acro{MEC}{mobile edge computing}
\acro{MSE}{mean square error}
\acro{LLR}{log-likelihood ratio}
\acro{LO}{local oscillator}
\acro{LOS}{line-of-sight}
\acro{LMMSE}{linear minimum mean squared error}
\acro{LNA}{low noise amplifier}
\acro{LSF}{local scattering function}
\acro{LTE}{long term evolution}
\acro{LUT}{look-up table}
\acro{LTV}{linear time-variant }
\acro{MIMO}{multiple-input multiple-output}
\acro{MPC}{multi-path component}
\acro{MC}{Monte Carlo}
\acro{NI}{National Instruments}
\acro{LoS}{line-of-sight}
\acro{NLOS}{non-line of sight}
\acro{OFDM}{orthogonal frequency division multiplexing}
\acro{OTA}{over-the-air}
\acro{PA}{power amplifier}
\acro{PC}{personal computer}
\acro{PDP}{power delay profile}
\acro{PER}{packet error rate}
\acro{PPS}{pulse per second}
\acro{QAM}{quadrature ampltiude modulation}
\acro{QPSK}{quadrature phase shift keying}

\acro{RB}{resource block}
\acro{RBP}{resource block pair}
\acro{RIS}{reflective intelligent surface}
\acro{RF}{radio frequency}
\acro{RMS}{root mean square}
\acro{RSSI}{receive signal strength indicator}
\acro{RT}{ray tracing}
\acro{RX}{receiver}
\acro{SCME}{spatial channel model extended}
\acro{SDR}{software defined radio}
\acro{SISO}{single-input single-output}
\acro{SoCE}{sum of complex exponentials}
\acro{SNR}{signal-to-noise ratio}
\acro{SUT}{system-under test}
\acro{SSD}{soft sphere decoder}
\acro{TBWP}{time-bandwidth product}
\acro{TDL}{tap delay line}
\acro{TX}{transmitter}
\acro{UMTS}{universal mobile telecommunications systems}
\acro{UDP}{user datagram protocol}
\acro{URLLC}{ultra-reliable and low latency communication} 
\acro{US}{uncorrelated-scattering}
\acro{USRP}{universal software radio peripheral}
\acro{VNA}{vector network analyzer}
\acro{ViL}{vehicle-in-the-loop}
\acro{V2I}{vehicle-to-infrastructure}
\acro{V2V}{vehicle-to-vehicle}
\acro{V2X}{vehicle-to-everything}
\acro{VST}{vector signal transceiver}
\acro{VTD}{Virtual Test Drive}
\acro{WF}{Wiener filter}
\acro{WSS}{wide-sense-stationary}
\acro{WSSUS}{wide-sense-stationary uncorrelated-scattering}
\end{acronym}

\begin{abstract}
Reconfigurable intelligent surfaces (RISs) are emerging as key enablers of reliable industrial automation in the millimeter-wave (mmWave) band, particularly in environments with frequent line-of-sight (LoS) blockage. While prior works have largely focused on theoretical aspects, real-time validation under user mobility remains underexplored. In this work, we propose and experimentally evaluate an adaptive beamforming algorithm that enables RIS reconfiguration via a low-rate feedback link from the mobile user equipment (UE) to the RIS controller, operating without requiring UE position knowledge. The algorithm maintains the received signal power above a predefined threshold using only a single-bit comparison of received power levels. To analyze the algorithm's performance, we establish a simulation-based Monte Carlo (MC) optimization benchmark that assumes full UE position knowledge, accounts for practical hardware constraints, and serves as an upper bound for performance evaluation. Using a hexagonal RIS with 127 elements and 1-bit phase quantization at 23.8\,GHz, we validate the proposed approach in a semi-anechoic environment over a 60\,cm$\times$92\,cm area. The results demonstrate that the single-bit feedback-driven algorithm closes much of the performance gap to the MC upper bound while achieving up to 24\,dB gain in received power compared to an inactive RIS baseline. These findings highlight the practical potential of feedback-based adaptive RIS control for robust mmWave non-line-of-sight (NLoS) communication with mobile users.
\end{abstract}


\section{Introduction}
Reconfigurable intelligent surfaces (RISs) have shown promising capabilities for enhancing indoor wireless systems, particularly in the millimeter-wave (mmWave) band \cite{Tang21, Tang22}. However, most of the existing works assume perfect channel state information (CSI) or accurate knowledge of the user equipment (UE) position, which are assumptions that are impractical in dynamic or mobile environments. Therefore, real-time channel estimation and RIS reconfiguration are essential for mobile applications and dynamic scenarios. We showed that as soon as the UE moves, it can experience a significant decay in received power \cite{radpour_reconfigurable_2024}. 

One approach in literature is to repeatedly transmit pilot signals from the receiver and measure the received signal strength using different RIS configurations. Therefore, a control loop is required between the UE and the transmitter or the RIS. For instance, \cite{zheng_intelligent_2020} proposes a feedback-driven RIS optimization method assuming a direct UE-BS link, which may not always be available in NLoS scenarios.

The basic idea of controlling a RIS using a UE-RIS feedback link is explored in \cite{Bjornson20a}. The authors of \cite{pei_ris-aided_2021-1} investigate a similar idea using a UE-RIS feedback link and propose an iterative algorithm to control the RIS reflection coefficients. The authors of \cite{pei_ris-aided_2021-1} apply this technique to provide measurement results in the sub-6 GHz frequency band for indoor and outdoor scenarios for a stationary UE. 

Another approach is embedding some active sensing elements in the RIS hardware structure to facilitate channel estimation. In this case, again, a control loop is required for a continuous CSI acquisition and updating of the RIS configuration in real-time. A deep learning study using active sensing elements is discussed in \cite{sohrabi_active_2022,jiang_active_2023,hwang_environment-adaptive_2023}, though these solutions require complex hardware integrations and channel models.

In this paper, we propose a practical and adaptive beamforming algorithm for RIS-assisted communication in dynamic NLoS environments. Our approach enables reliable communication with a moving UE, without requiring explicit knowledge of the UE’s location. The algorithm operates via a low-rate feedback link from the UE to the RIS controller, which adjusts the RIS configuration in real time based on signal strength. The proposed method supports general RIS structures and is validated experimentally using a hexagonal RIS prototype in a controlled testbed environment.

\subsection*{Scientific Contributions}
\begin{itemize}
    \item We propose an adaptive RIS control algorithm that updates configurations in real time using a low-rate single-bit feedback link, operating without UE position knowledge to maintain reliable mmWave NLoS links.
    \item We establish a Monte Carlo (MC) optimization benchmark with full UE position knowledge, based on our previous work \cite{radpour_reconfigurable_2024,radpour_active_2024}, providing a ground-truth upper bound to evaluate the proposed feedback-driven algorithm.
    \item We generalize the iterative RIS control framework from \cite{pei_ris-aided_2021-1} to support more complex RIS layouts, including hexagonal structures, and extend it to dynamic environments with mobile UEs, broadening the applicability of feedback-based RIS control strategies.
    \item We develop and validate a signal model and conduct detailed measurements in a semi-anechoic room, demonstrating up to 24\,dB received power gain and providing practical insights into RIS-assisted communication under mobility.
\end{itemize}

\section{Signal Model}
\label{sigmodel}
\begin{figure}[ht!]
	\centering
	\includegraphics[width=\columnwidth]{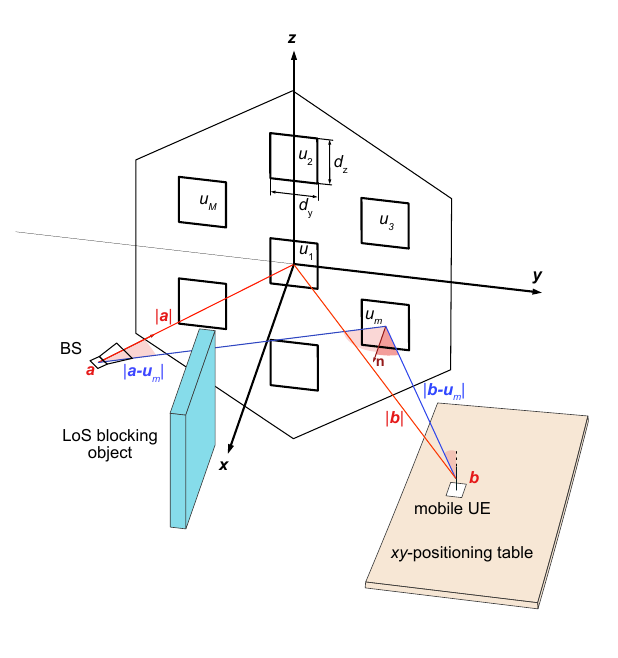}
	\caption{RIS coordinate system for a hexagonal RIS element placement in the yz-plane. The BS horn antenna radiates from position $\Vec{a}$ towards the center of the RIS at $\Vec{0}=(0,0,0)$ over a distance of $|\Vec{a}|$, similarly the UE omni-directional antenna at position $\Vec{b}$ is within a distance of $|\Vec{b}|$. The LOS is blocked between BS and UE. The picture is not to scale to improve clarity.} 
	\label{fig:RIScoordinates}
\end{figure}

In this section, we discuss the signal model that is used to calculate the received signal power at the UE for the numerical simulation. We consider that the UE is in NLoS of the transmitter, and the transmitter and the UE are in direct LoS of the RIS; please see Fig.~\ref{fig:RIScoordinates} for further details \cite{radpour_reconfigurable_2024,radpour_active_2024}. The RIS element $m$ can have the complex reflection coefficient $\Gamma_m \in \mathcal{A}$, where $\mathcal{A}$ is the alphabet of all the possible reflection coefficients $\gamma_k$, and $K$ is the number of possible RIS element phase quantizations:
\[
\mathcal{A}=\{\gamma_1, \ldots, \gamma_K\}.
\]
The received signal $y$ at the UE, reflected from the RIS can be stated as
\begin{equation}
y=h s+n  
\end{equation}
where $s$ is the transmitted signal, $ n \sim C \mathcal{N}\left(0, \sigma^2\right)$ is the additive white Guassian noise and  $h$ is the frequency flat channel coefficient \cite{Tang21, hu_reconfigurable_2020}
\begin{multline}
h=\frac{\sqrt{G_\text{BS} G_\text{UE}}(d_y d_z)}{4\pi}
\times 
\sum_{m=1}^{M} \Gamma_m \frac{\sqrt{F^\text{c}_m}e^{-j 2 \pi\left(|\Vec{a}-\Vec{u}_m|+|\Vec{b}-\Vec{u}_m|\right)/\lambda}}{(|\Vec{a}-\Vec{u}_m| |\Vec{b}-\Vec{u}_m|)}.
\end{multline}
and the received signal power at the UE can be stated as 
\begin{equation}
    P_{\text{UE}}=P_{\text{BS}}\left | h\right |^2+P_\text{N}
\end{equation} 
where $P_{\text{BS}}$, $P_\text{N}$, $G_{\text{BS}}$, $G_{\text{UE}}$, $d_\text{y}$, $d_\text{z}$ and $M$ are the BS transmit power, the noise power, the BS and UE antenna gain, the effective RIS element dimension in $y$ and $z$ direction and the number of RIS elements, respectively \cite{Tang21, Tang22, radpour_reconfigurable_2024}. The wavelength $\lambda=c_0/f$, where $f$ denotes the center frequency and $c_0$ is the speed of light. The combined antenna pattern of the BS antenna, the RIS element $m$ for receive and transmit operation, as well as the UE antenna is described by $F^\text{c}_m$, please see \cite[(2)]{radpour_reconfigurable_2024} for details.

For our RIS working in the active mode \cite{radpour_reconfigurable_2024} at the center frequency of 23.8 GHz, we have a 1-bit reflection coefficient alphabet, resulting in $K=2$ states
\[
\mathcal{A}=\{(1.25, \angle 0^\circ), ( 0, \angle 0^\circ) \}.
\] 

\RestyleAlgo{ruled}
\begin{algorithm}
\caption{ Adaptive beamforming algorithm}
\label{alg:Iteralg}
\KwIn{ Feedback of the received signal power $P_\text{UE}$, power threshold $P_\text{TS}$ and maximum iteration size $\mathcal{I}$.}
\KwOut{Reflection coefficient matrix $\mathcal{R}^{(I)}$, received power after all iterations $P_\text{UE}^{(I)}$.}

\ Initialize a random ref. coef. matrix $\mathcal{R}^{(0)}$\;
 \If{$P_\textup{UE} < P_\textup{TS}$}{
 \ \text{Feedback to the RIS controller}\;
\For{each $i\in [1,\mathcal{I}]$}{
    \For{each $j \in [1,\mathcal{J}]$}{
      \For{each $l \in [1,\mathcal{L}^j]$}{
        \For{each \textup{RIS element} $m \in \mathcal{G}^j_l$}{ 
 Update reflection coefficient: $\gamma_k \gets \text{Switch}(\gamma_k)$\;                                        }
           \If{$P_\textup{UE}^{(i)} < P_\textup{UE}^{(i-1)}$}{
            \ \text{Feedback to the RIS controller}\;
             $\quad \mathcal{R}^{(i)} \gets \mathcal{R}^{(i-1)}$\;
                                                      }
               \ $i \gets i+1$\;
               \If{$i > \mathcal{I}$}{\textbf{break}}
                            } } }}
\end{algorithm}

\section{Adaptive Algorithm for a Mobile UE}
An iterative beamforming algorithm based on a rectangular RIS is introduced in \cite{pei_ris-aided_2021-1}, targeting stationary UEs. In contrast, our work extends this concept to flexible RIS structures, supports both reflective and active modes, and considers a mobile UE in a NLoS setting. We propose an algorithm that generalizes the control strategy to arbitrary RIS structures and leverages a low-rate feedback link from the UE to the RIS controller. This feedback enables the RIS to adapt its configuration in real-time to maintain the received signal power above a pre-defined threshold $P_\text{TS}$, which is a value determined by the UE’s sensitivity. Notably, this approach does not require knowledge of the UE's position.

The RIS elements are organized into subgroups based on the geometry of the RIS surface. To ensure smooth and stable adaptation, the grouping strategy must minimize significant power fluctuations and allow for overlap across iterations. We define $\mathcal{J}$ group sets, where each group set $\mathcal{G}^j$, for $j \in [1, \mathcal{J}]$, consists of $\mathcal{L}^j$ subgroups, with $\mathcal{L}^j = |G^j|$, where $\left| \cdot \right|$ denotes the cardinality, representing the number of elements in a set. Accordingly, we can express the grouping structure as:
\[
\mathcal{G} = \left\{ \mathcal{G}^1, \mathcal{G}^2, \ldots \mathcal{G}^\mathcal{J} \right\}.
\]
and for each $j$-th group, we define:
\[
\mathcal{G}^j = \left\{ \mathcal{G}^j_1, \mathcal{G}^j_2, \ldots, \mathcal{G}^j_{\mathcal{L}^j} \right\}.
\]
Here, $\mathcal{G}$ represents the collection of all $\mathcal{J}$ group sets, and $\mathcal{G}^j_l$ refers to the $l$-th subgroup of the $j$-th group, containing the indices of all RIS elements associated with that subgroup.

\begin{figure}
	\centering
    {\includegraphics[width=\columnwidth]{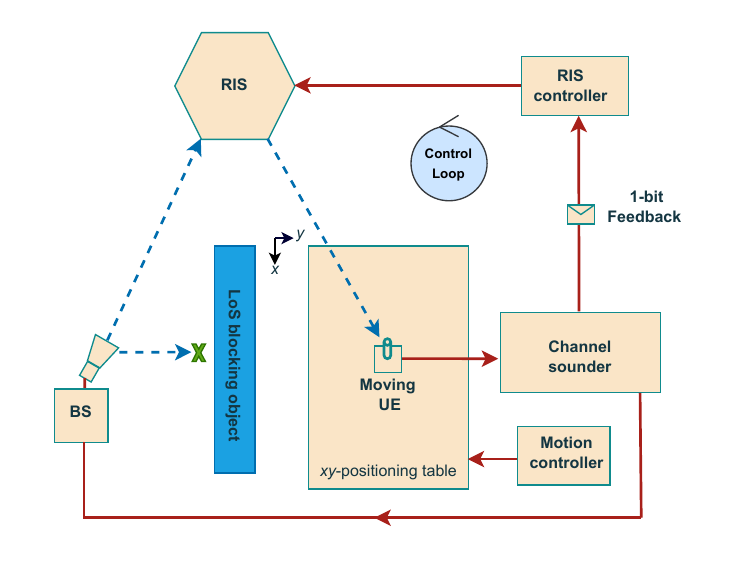}}
	\caption{Block diagram of the measurement setup used to evaluate the RIS-aided iterative beamforming algorithm. The feedback link between the UE and the RIS controller forms a closed-loop control system involving the UE, the RIS controller, and the RIS. The UE moves inside the area of interest (AoI), which is in NLoS of BS.}
	\vspace{-4mm}
    \label{fig:blockdiagram}
\end{figure}

Algorithm~\ref{alg:Iteralg} continuously monitors the received power at the UE. If the received power falls below the threshold $P_\text{TS}$, the algorithm iteratively tests a new RIS configuration by flipping the state of a sub-group. If the new configuration results in an improved received power, it is retained; otherwise, it is reverted. The reflection coefficient update for each RIS element is controlled by the Switch($\cdot$) function, which toggles between the values $\gamma_1$ and $\gamma_2$. The testbed setup is illustrated in Figure~\ref{fig:blockdiagram}, and the complete configuration update process is outlined in Algorithm~\ref{alg:Iteralg}.

In each iteration $i$, Algorithm~\ref{alg:Iteralg} systematically switches the reflection coefficient of each RIS element $m$ within the $l$-th subgroup of the $j$-th group. The comparison of the received power between consecutive iterations, $P_\text{UE}^{(i)} < P_\text{UE}^{(i-1)}$, is signaled back to the RIS controller via a single-bit feedback link. Based on this single-bit feedback, the RIS controller determines whether to retain the current RIS configuration $\mathcal{R}^{(i)}$ or revert to the previous configuration $\mathcal{R}^{(i-1)}$. The algorithm is designed to run for a maximum of $\mathcal{I}$ iterations to assess the received power behavior over all iterations, though it can be modified to terminate early if $P_\text{UE} > P_\text{TS}$. This compact signaling mechanism reduces feedback overhead and enables practical real-time RIS adaptation in mobile NLoS scenarios, without requiring full CSI or high-rate feedback.

\begin{figure}
	\centering
	\includegraphics[width=\columnwidth]{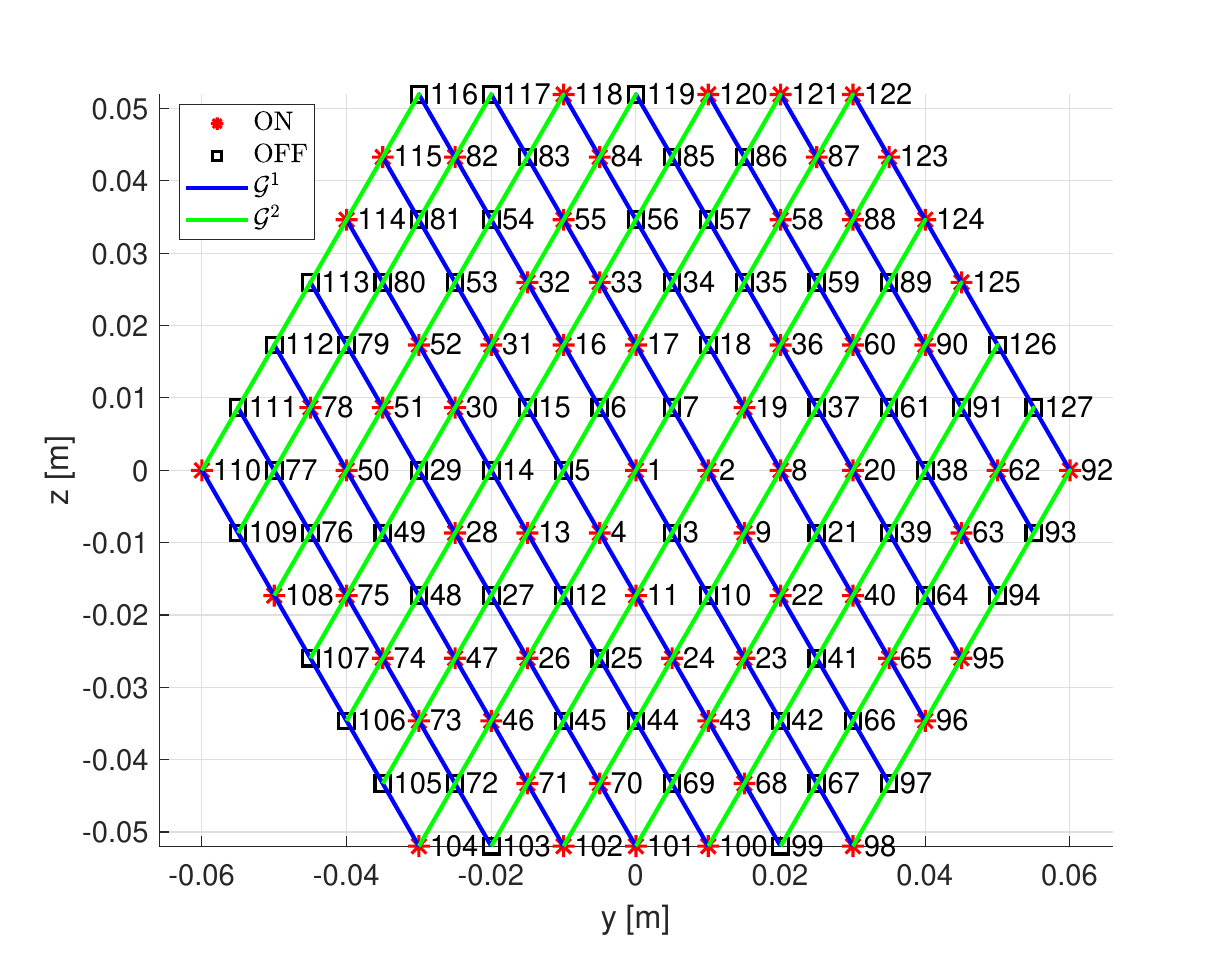} 
	\caption{RIS element grouping structure for the proposed iterative algorithm. Blue and green boundary lines correspond to the subgroups of $\mathcal{G}^1$ and $\mathcal{G}^2$, respectively. For element indices, please refer to Tables~\ref{tab:group1} and \ref{tab:group2}.}
	\label{fig:RISGrouping}
\end{figure}

While \cite{pei_ris-aided_2021-1} employed row- and column-based flipping for a rectangular RIS, our hexagonal structure requires a different strategy. We define two directional groups ($\mathcal{J}=2$), $\mathcal{G}^1$ and $\mathcal{G}^2$, each consisting of 13 subgroups ($\mathcal{L}^1=\mathcal{L}^2=13$), slanted at $\pm 30^\circ$ to align the physical layout of the RIS.  The grouping layout is illustrated in Figure~\ref{fig:RISGrouping}, while Tables~\ref{tab:group1} and \ref{tab:group2} provide the element indices for each subgroup. Our RIS features a 1-bit reflection coefficient alphabet $\mathcal{A} = \{\gamma_1, \gamma_2\}$, enabling each element to toggle between two possible states.

For benchmarking, we also evaluate the Monte Carlo optimization approach previously introduced in \cite{radpour_reconfigurable_2024,radpour_active_2024}, which assumes full knowledge of the UE position. This benchmark also accounts for practical RIS constraints, including hardware impairments and quantized (1-bit) phase control. To ensure a fair comparison, both the proposed iterative single-bit feedback algorithm and the MC approach are compared after 100 iterations.

\begin{table}[htbp]
\caption{Group 1 ($\mathcal{G}^{1}$)}
\label{tab:group1}
\centering
\renewcommand{\arraystretch}{1.75} 
\begin{tabular}{|c|l|}
\hline
\textbf{Sub-Group} & \textbf{RIS Element Indice} \\
\hline
$\mathcal{G}^{1}_{1}$  & 104, 105, 106, 107, 108, 109, 110 \\
$\mathcal{G}^{1}_{2}$  & 103, 72, 73, 74, 75, 76, 77, 111 \\
$\mathcal{G}^{1}_{3}$  & 102, 71, 46, 47, 48, 49, 50, 78, 112 \\
$\mathcal{G}^{1}_{4}$  & 101, 70, 45, 26, 27, 28, 29, 51, 79, 113 \\
$\mathcal{G}^{1}_{5}$  & 100, 69, 44, 25, 12, 13, 14, 30, 52, 80, 114 \\
$\mathcal{G}^{1}_{6}$  & 99, 68, 43, 24, 11, 4, 5, 15, 31, 53, 81, 115 \\
$\mathcal{G}^{1}_{7}$  & 98, 67, 42, 23, 10, 3, 1, 6, 16, 32, 54, 82, 116 \\
$\mathcal{G}^{1}_{8}$  & 97, 66, 41, 22, 9, 2, 7, 17, 33, 55, 83, 117 \\
$\mathcal{G}^{1}_{9}$  & 96, 65, 40, 21, 8, 19, 18, 34, 56, 84, 118 \\
$\mathcal{G}^{1}_{10}$ & 95, 64, 39, 20, 37, 36, 35, 57, 85, 119 \\
$\mathcal{G}^{1}_{11}$ & 94, 63, 38, 61, 60, 59, 58, 86, 120 \\
$\mathcal{G}^{1}_{12}$ & 93, 62, 91, 90, 89, 88, 87, 121 \\
$\mathcal{G}^{1}_{13}$ & 92, 127, 126, 125, 124, 123, 122 \\
\hline
\end{tabular}
\end{table}

\begin{table}[htbp]
\caption{Group 2 ($\mathcal{G}^{2}$)}
\label{tab:group2}
\centering
\renewcommand{\arraystretch}{1.75} 
\begin{tabular}{|c|l|}
\hline
\textbf{Sub-Group} & \textbf{RIS Element Indice} \\
\hline
$\mathcal{G}^{2}_{1}$  & 98, 97, 96, 95, 94, 93, 92 \\
$\mathcal{G}^{2}_{2}$  & 99, 67, 66, 65, 64, 63, 62, 127 \\
$\mathcal{G}^{2}_{3}$  & 100, 68, 42, 41, 40, 39, 38, 91, 126 \\
$\mathcal{G}^{2}_{4}$  & 101, 69, 43, 23, 22, 21, 20, 61, 90, 125 \\
$\mathcal{G}^{2}_{5}$  & 102, 70, 44, 24, 10, 9, 8, 37, 60, 89, 124 \\
$\mathcal{G}^{2}_{6}$  & 103, 71, 45, 25, 11, 3, 2, 19, 36, 59, 88, 123 \\
$\mathcal{G}^{2}_{7}$  & 104, 72, 46, 26, 12, 4, 1, 7, 18, 35, 58, 87, 122 \\
$\mathcal{G}^{2}_{8}$  & 105, 73, 47, 27, 13, 5, 6, 17, 34, 57, 86, 121 \\
$\mathcal{G}^{2}_{9}$  & 106, 74, 48, 28, 14, 15, 16, 33, 56, 85, 120 \\
$\mathcal{G}^{2}_{10}$ & 107, 75, 49, 29, 30, 31, 32, 55, 84, 119 \\
$\mathcal{G}^{2}_{11}$ & 108, 76, 50, 51, 52, 53, 54, 83, 118 \\
$\mathcal{G}^{2}_{12}$ & 109, 77, 78, 79, 80, 81, 82, 117 \\
$\mathcal{G}^{2}_{13}$ & 110, 111, 112, 113, 114, 115, 116 \\
\hline
\end{tabular}
\end{table}

\begin{figure}
	\centering
    {\includegraphics[width=\columnwidth]{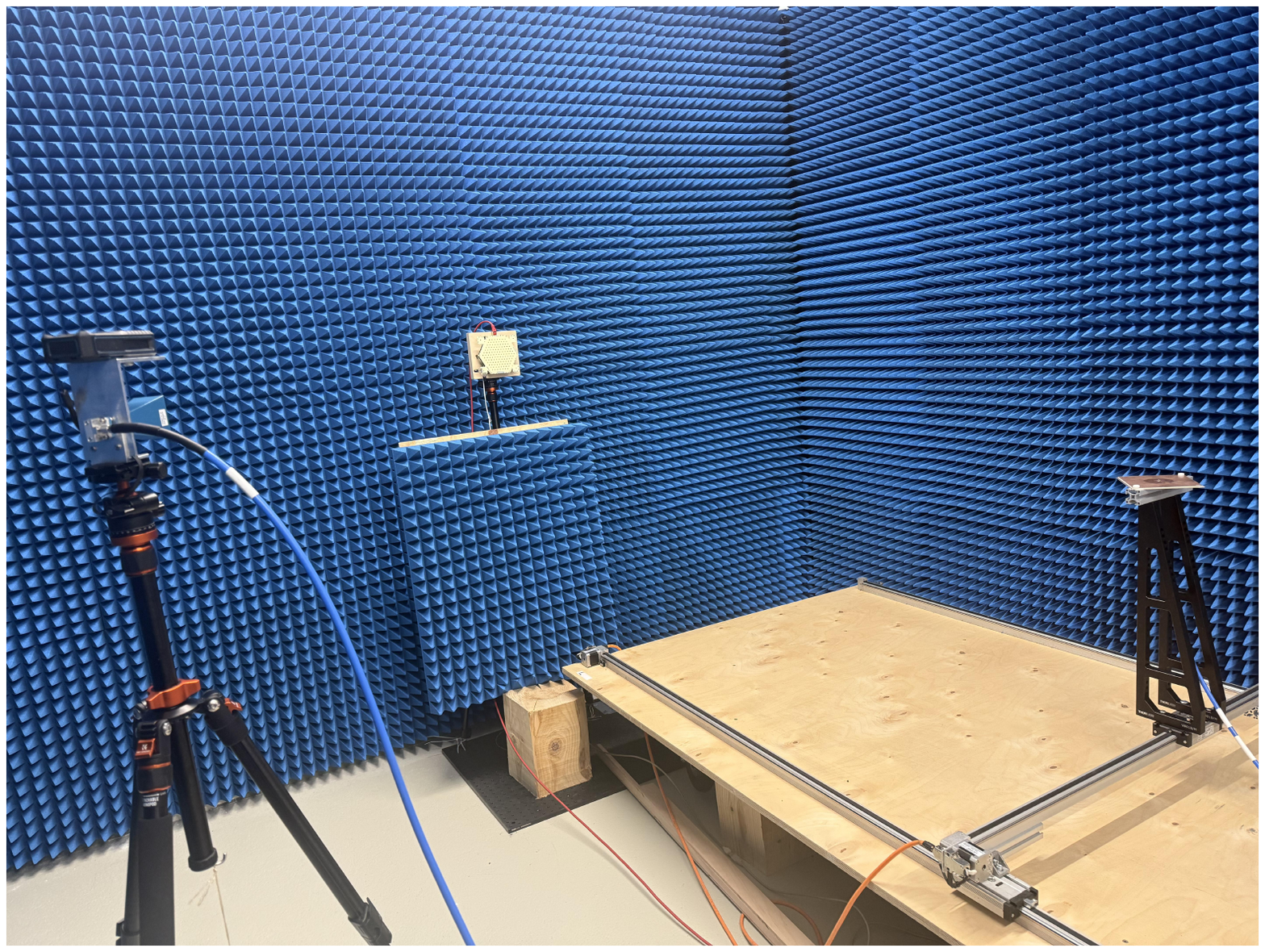}}
	\caption{Photograph of the RIS measurement testbed (LoS absorber removed for clarity). The horn antenna serves as the transmitter, and the UE moves on an $xy$-positioning table equipped with an omnidirectional monopole antenna.}
	\vspace{-4mm}
    \label{fig:testbed}
\end{figure}

\begin{table}
\begin{center}
\caption{Active RIS Parameters and Measurement Setup.}
\label{tab:systemparam}
\begin{tabular}{ll} 
\toprule
Parameter	      	&  Definition\\
\midrule
$f = 23.8$\,GHz		& 	 center frequency \\
$M = 127$    & number of RIS elements\\
$d_z$, $d_y = 6.6$\,mm	& 	 effective RIS element size \\
$d = 8.7$\,mm & smallest RIS element distance\\
\midrule
$P_\text{BS}=10$\,dBm		& 	 BS transmit power\\
$G_\text{BS} = 19$\,dB		&  BS horn antenna gain\\
$G_\text{UE} = 3.2$\,dB		&   UE vert. pol. monopole antenna gain\\
$|\Vec{a}|= 1.86$\,m  & RIS-BS distance\\
${\Vec{a}} =( 1.5\,\text{m}, -1.09\,\text{m}, 0)$ & BS location\\
\bottomrule
\end{tabular}
\end{center}
\end{table}

\begin{figure}
	\centering
    {\includegraphics[width=\columnwidth]{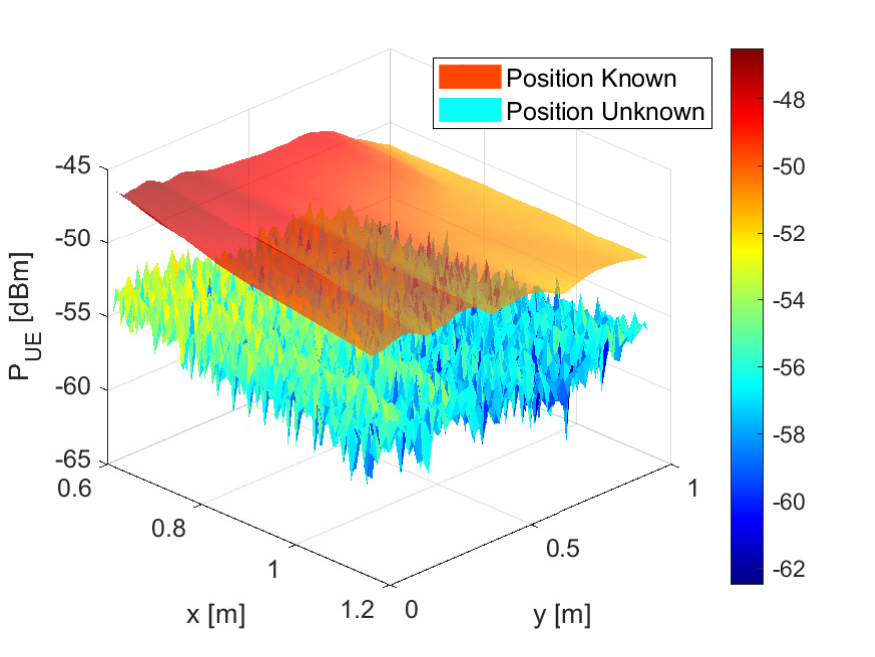}} 
	\caption{Simulated received power distribution over the area of interest for two RIS control strategies: (i) Monte Carlo (MC) optimization with full UE position knowledge (upper-bound), and (ii) proposed iterative single-bit feedback algorithm without position knowledge. This simulation provides a benchmark for assessing the performance gap between the ideal MC optimization and the proposed practical feedback-driven approach.}
	\vspace{-4mm}
    \label{fig:MCvsIter}
\end{figure}

\begin{figure*}
	\centering
	\subfloat[Meas. result for $P_\text{TS}=-70\,\text{dBm}$.]
 {\includegraphics[width=\columnwidth]{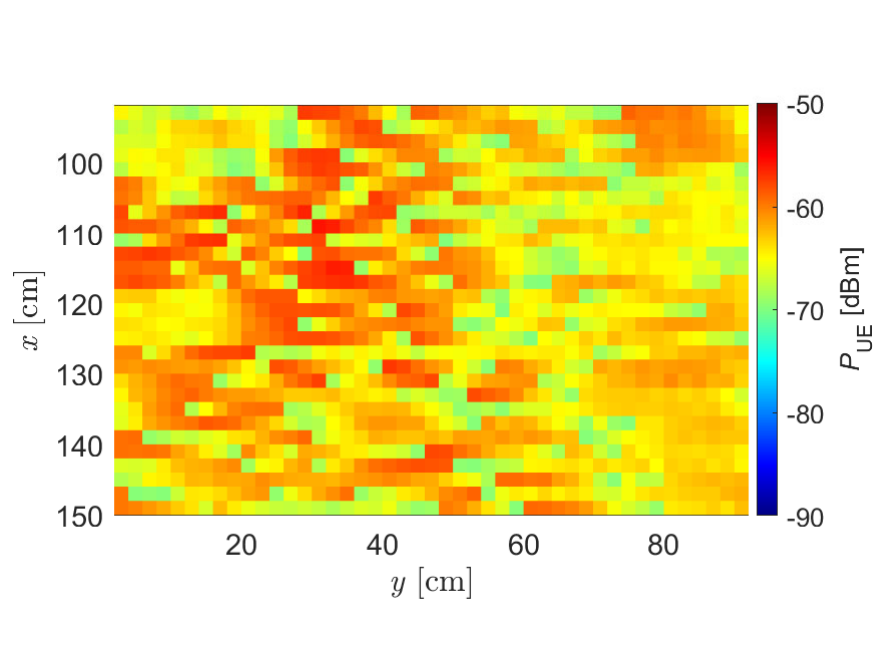} \label{fig:Pts-70}}
 	\subfloat[Meas. result for $P_\text{TS}=-65\,\text{dBm}$.]{\includegraphics[width=\columnwidth]{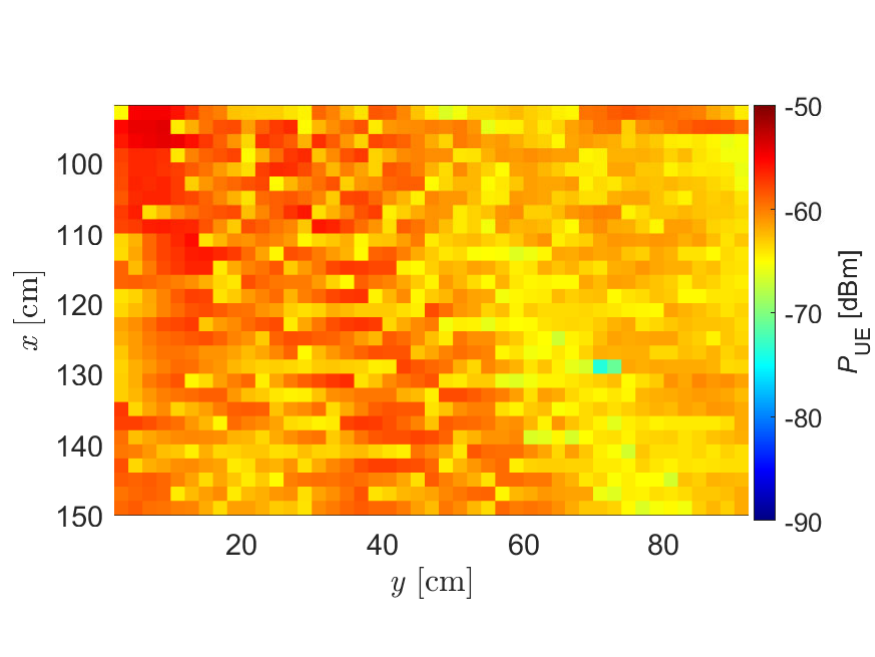}\label{fig:Pts-65}}
    \caption{Received power $P_\text{UE}(x, y)$ over observation area. We compare the empirical measurement results using the iterative algorithm with 100 iterations for two different threshold powers of (a) $P_\text{TS}=-70\,\text{dBm}$, (b) $P_\text{TS}=-65\,\text{dBm}$. The BS is located at ${\Vec{a}} =( 1.5\,\text{m}, -1.09\,\text{m}, 0)$, and the UE is moving through all the observation points of the $xy$-positioning table.}
	\label{fig:2MeasResult}
		\vspace{-4mm}
\end{figure*}

\begin{figure}
	\centering
	\includegraphics[width=\columnwidth]{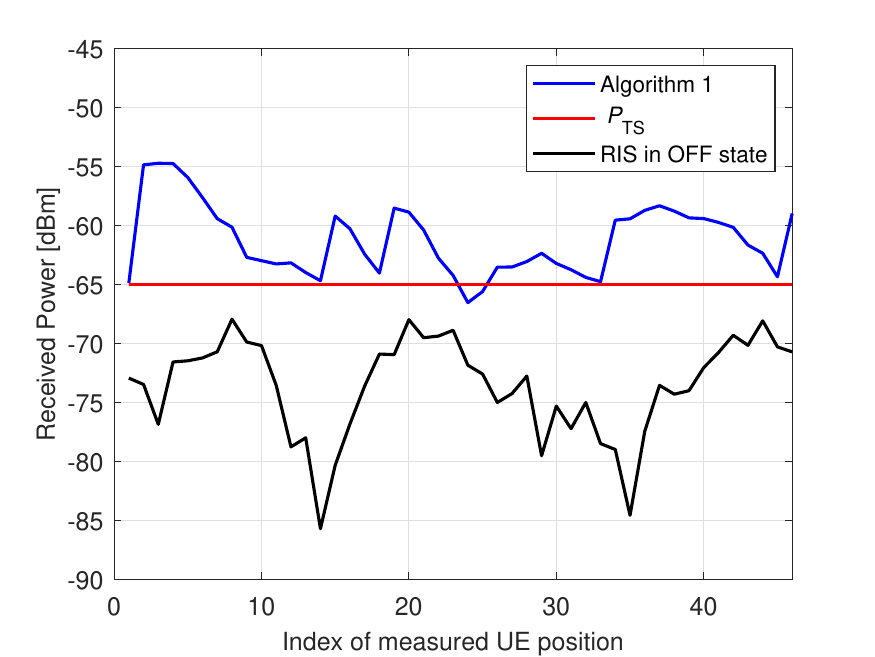} 
    \caption{Comparison of the received power over all the measured points along a single measurement row at $x = 92\,\text{cm}$ and $y \in \{0, 2, 4, \ldots, 90\}\,\text{cm}$ for the cases (i) RIS in the off state, and (ii) RIS controlled by Algorithm~1.}
	\label{fig:BeforeAfterComparison}
\vspace{-4mm}
\end{figure}

\begin{figure}
	\centering
	\includegraphics[width=\columnwidth]{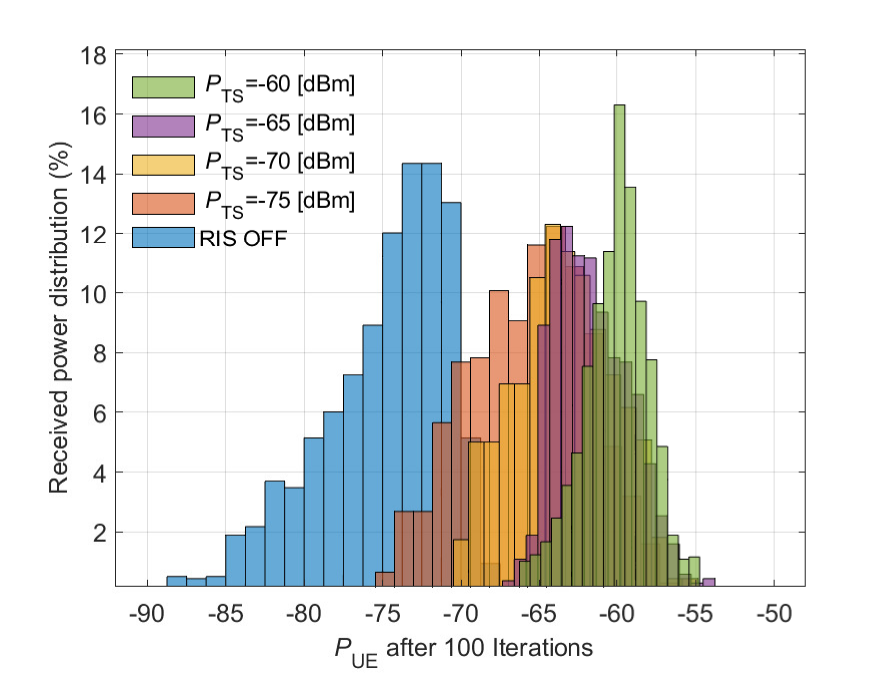} 
	\caption{Histogram of the measurement results for the case that all RIS elements are turned off (RIS OFF) and four different $P_{\text{TS}}$ after 100 iterations.} 
	\label{fig:histogram}
\vspace{-4mm}
\end{figure}

\section{Simulation and Measurement Results}
In this section, we validate the performance of the proposed Algorithm~\ref{alg:Iteralg} through both simulations and real-time measurements using our testbed. First, we simulate the algorithm based on the signal model introduced in Section \ref{sigmodel}, followed by its implementation for real-time measurement. We conduct the measurements using AIT's mmWave channel sounder \cite{molisch_millimeter-wave_2025}, and record the channel impulse responses (CIRs). Table~\ref{tab:systemparam} summarizes the key system parameters, Fig.~\ref{fig:blockdiagram} shows the block diagram of the measurement setup and Fig.~\ref{fig:testbed} illustrates the experimental setup.

The mobile UE is mounted on an $xy$-positioning table, which enables precise movement across a defined area of interest in the $xy$-plane spanning $60\,\text{cm} \times 92\,\text{cm}$. Signal measurements are recorded every 2\,cm within this AoI.
These measurements are fed back to the RIS controller using a user datagram protocol (UDP) connection. To calculate the received power at each UE location, we perform five repeated measurements to record the channel sounder data. We then average the recorded CIRs and compute the received power using Eq.~(2) in \cite{radpour_reconfigurable_2024}.
 To ensure a fair comparison, we set the maximum number of iterations to $\mathcal{I} = 100$ for both the simulation and measurement experiments in this section.

\subsection{Simulated Comparison: Iterative vs. Monte Carlo}
Figure~\ref{fig:MCvsIter} presents a simulation-based analysis comparing the proposed iterative single-bit feedback algorithm with the Monte Carlo (MC) optimization method \cite{radpour_reconfigurable_2024,radpour_active_2024}. The MC approach assumes full UE position knowledge and serves as an upper-bound benchmark for RIS control. In contrast, the iterative algorithm operates without position information, relying only on a low-rate single-bit feedback link. Both methods are evaluated after 100 iterations to ensure a fair comparison of their performance. The results show that while the MC approach achieves higher received power across the area of interest, the iterative feedback-driven algorithm closes much of this gap, demonstrating that adding a low-rate feedback link effectively compensates for the lack of position knowledge and enables NLoS mmWave communication.

\subsection{Measured Results: Received Power vs. Position}
Figure~\ref{fig:2MeasResult} demonstrates the received power pattern using Algorithm~1 for a moving UE over the AoI, which spans $60 \times 92$\,cm. The AoI is sampled at 2\,cm intervals, resulting in 30 × 46 = 1380 unique UE positions. Figures~\ref{fig:Pts-70} and \ref{fig:Pts-65} show the received power for two different power thresholds $P_\text{TS} \in \{-70\,\text{dBm},-65\,\text{dBm}\}$. A higher threshold leads to increased average power, showing the algorithm's adaptability. The observed pattern consists of gradual changes followed by sharp power jumps, reflecting the algorithm's trigger-based adaptation. This sudden increase is a result of the trigger-based nature of the algorithm, which only engages if the received power of the UE falls below the specified threshold $ P_{\text{TS}}$ and otherwise maintains the current RIS configuration and moves to the next step.

Figure~\ref{fig:BeforeAfterComparison} compares the received power with and without RIS activation along a single measurement row at $x = 92\,\text{cm}$ and $y \in \{0, 2, 4, \ldots, 90\}\,\text{cm}$.
 With the RIS in the off state, the elements act as passive reflectors with no amplification or phase control. Figure~\ref{fig:BeforeAfterComparison} illustrates that using the iterative algorithm results in an improvement of up to 24\,dB gain in the received power levels. Nevertheless, it is noted that the received power at some points falls below the pre-defined threshold. This is because the algorithm detailed in Algorithm~\ref{alg:Iteralg} can increase power; however, it does not consistently ensure that outputs exceed a specified threshold. Furthermore, it is essential to highlight that increasing the threshold power generally leads to higher received power levels, yet more points may fall below the threshold line.

Finally, Fig.~\ref{fig:histogram} presents histograms of received power for different values of $P_\text{TS}\in\{-60, -65, -70, -75\}\,\text{dBm}$, along with the baseline where the RIS is in the off state. Results confirm that applying the iterative algorithm consistently boosts received power across the chosen observation area. Moreover, increasing the threshold results in overall higher power levels, at the cost of some points falling below the set threshold.

\section{Conclusion}
In this paper, we present a generalized adaptive beamforming algorithm for RIS-assisted mmWave communication in mobile NLoS scenarios. The proposed method ensures reliable signal reception by leveraging a low-rate single-bit feedback mechanism and supporting flexible RIS structures. To provide a benchmark, we establish a simulation-based Monte Carlo optimization approach assuming full UE position knowledge as an upper bound. Experimental validation using a 127-element hexagonal RIS demonstrates that the proposed algorithm closes much of the performance gap to the MC benchmark while achieving up to 24\,dB improvement in received power for a mobile UE, confirming its effectiveness in real-world settings. This work demonstrates that effective RIS control is achievable with minimal feedback overhead, enabling scalable and practical deployment in 6G-enabled industrial and mobile communication systems.

\section*{Acknowledgment}
This work is funded by the Principal Scientist grant at the AIT Austrian Institute of Technology within the project DEDICATE. We would like to thank Mike Hentschel for his support in the photonic identification of the 3D measurement geometry.

\bibliographystyle{IEEEtran}
\bibliography{IEEEabrv, DissLib, references}

\end{document}